\documentclass[sigconf,natbib]{acmart}
\usepackage{url}
\usepackage{multirow}
\usepackage{float}
\usepackage{balance}
\usepackage{color}
\usepackage{graphics,graphicx}
\usepackage{soul}
\usepackage{subcaption}
\usepackage{natbib}
\usepackage{enumitem}
\usepackage[justification=centering]{caption}

\newcommand{\ift}[1][abbr]{%
  \ifthenelse{\equal{#1}{full}}{Information Foraging Theory}{%
  \ifthenelse{\equal{#1}{first}}{Information Foraging Theory (IFT)}
  \ifthenelse{\equal{#1}{short}}{Information Foraging}{%
  IFT}}%
}

\newcommand{\bpg}[1][abbr]{%
  \ifthenelse{\equal{#1}{full}}{Berry Picking}{%
  \ifthenelse{\equal{#1}{first}}{Berrying Picking (BP)}{%
  BP~}}%
}

\newcommand{\ifg}[1][abbr]{%
  \ifthenelse{\equal{#1}{full}}{Information Farming}{%
  \ifthenelse{\equal{#1}{first}}{Information Farming (IF)}{%
  IF~}}%
}

\newcommand{\genai}[1][abbr]{%
  \ifthenelse{\equal{#1}{full}}{Generative AI}{%
  \ifthenelse{\equal{#1}{first}}{Generative AI (GenAI)}{%
  GenAI~}}%
}

\newcommand{\llm}[1][abbr]{%
  \ifthenelse{\equal{#1}{full}}{Large Language Models}{%
    \ifthenelse{\equal{#1}{first}}{Large Language Models (LLMs)}{%
      \ifthenelse{\equal{#1}{s}}{LLMs~}{%
        LLM~}%
    }%
  }%
}

\newcommand{\rag}[1][abbr]{%
  \ifthenelse{\equal{#1}{full}}{Retrieval Augmented Generation}{%
  \ifthenelse{\equal{#1}{first}}{Retrieval Augmented Generation (RAG))}
  \ifthenelse{\equal{#1}{s}}{RAGs~}{%
  RAG~}}%
}

\newcommand{\ir}[1][abbr]{%
  \ifthenelse{\equal{#1}{full}}{Information Retrieval}{%
  \ifthenelse{\equal{#1}{first}}{Information Retrieval (IR)}{%
  IR~}}%
}

\AtBeginDocument{%
  \providecommand\BibTeX{{%
    \normalfont B\kern-0.5em{\scshape i\kern-0.25em b}\kern-0.8em\TeX}}}

\acmDOI{10.1145/3786304.3787947}
\acmISBN{979-8-4007-2414-5/2026/03}
\setcopyright{acmcopyright}
\copyrightyear{2026}
\acmYear{2026}
\acmPrice{}
 \copyrightyear{2026} 
 \acmYear{2026} 
 \setcopyright{acmlicensed}\acmConference[CHIIR '26]{2026 ACM SIGIR Conference on Human Information Interaction and Retrieval}{March 22--26, 2026}{Seattle, WA, USA}
\acmBooktitle{2026 ACM SIGIR Conference on Human Information Interaction and Retrieval (CHIIR '26), March 22--26, 2026, Seattle, WA, USA}

\begin{document}

\title{Information Farming}
\subtitle{From Berry Picking to Berry Growing}

\author{Leif Azzopardi}
\orcid{0000-0002-6900-0557}
\email{leifos@acm.org}
\affiliation{%
  \institution{University of Strathclyde}
  \city{Glasgow}
  \country{United Kingdom}
  }

  \author{Adam Roegiest}
\orcid{0000-0003-1265-8881}
\email{adam@roegiest.com}
\affiliation{%
  \institution{Zuva}
  \city{Kitchener}
  \country{Canada}
  }

\begin{abstract}
The classic paradigms of \bpg[full] and \ift[full] have framed users as gatherers, opportunistically searching across distributed sources to satisfy evolving information needs.
However, the rise of \genai[first] is driving a fundamental transformation in how people produce, structure, and reuse information—one that these paradigms no longer fully capture.
This transformation is analogous to the Neolithic Revolution, when societies shifted from hunting and gathering to cultivation.
Generative technologies empower users to ``\textit{farm}'' information by planting seeds in the form of prompts, cultivating workflows over time, and harvesting richly structured, relevant yields within their own plots, rather than foraging across others people's patches.
In this perspectives paper, we introduce the notion of \textit{\ifg[full]} as a conceptual framework and argue that it represents a natural evolution in how people engage with information.
Drawing on historical analogy and empirical evidence, we examine the benefits and opportunities of information farming, its implications for design and evaluation, and the accompanying risks posed by this transition.
We hypothesize that as \genai technologies proliferate, cultivating information will increasingly supplant transient, patch-based foraging as a dominant mode of engagement, marking a broader shift in human-information interaction and its study.
\end{abstract}


\maketitle

\section{Introduction}\label{lbl_introduction}
For decades, the Berry Picking~\cite{Bates1989TheInterface} and Information Foraging Theory~\cite{Pirolli1999} paradigms have shaped our understanding of interactive information seeking and retrieval. 
These frameworks described how searchers navigated information landscapes, visiting patches and gathering ``berries'' opportunistically, while providing researchers with robust methods for designing systems, conducting experiments, and analyzing user behavior.
Taken together, they offered an intuitive lens for understanding how people search and discover information, particularly during the formative years of online access~\cite{chi2001info_scent}.
However, the advent of \genai[first] systems, often underpinned by \llm[first], is transforming how people interact with information and information retrieval systems~\cite{Rahmani2024LLM4Eval,trippas2025report,shah2022situating,shah2023searchtask}. 
Analogous to the Neolithic Revolution, in which hunter-gatherer societies transitioned to agriculture, \genai enables a shift from \textit{gathering} information to \textit{growing} information.
Crucially, this transformation alters the underlying cost structure of information seeking: effort shifts away from navigating external patches toward shaping information within the interaction itself.
As a result, information foragers are becoming information farmers. 

By \textit{planting seeds} (prompts), users can \textit{cultivate} and \textit{harvest} information with the help of \genai systems and tools. 
This is because \llm[s] provide an information-dense medium that can be further enriched through retrieval augmentation~\cite{lewis2020rag} and agentic patterns~\cite{yao2023react} allowing users to shape and refine the desired outputs in their own plot.
Rather than foraging in patches to find the relevant pieces of information (berries), users can \textit{sow and grow} the information (berries) that they desire. 
As a result, they can obtain higher yields (more berries) at a lower costs (without the need to travel between patches).
For example, consider a researcher tasked with writing an article on ``\textit{the laws of Information Retrieval}''.
In the pre-\genai era, this would require extensive foraging: issuing multiple queries, inspecting results, visiting articles, following citations, and synthesizing findings~\cite{Kuhlthau1991}.
In contrast, with \genai, the researcher can \textit{farm} information by seeding an Agentic IR system 
to produce a tailored response. The researcher can then curate, prune and weed the output to harvest structured, relevant information in their desired format.

This shift is reshaping how people interact with retrieval systems, how information is produced and consumed, and the effort, structure, and quality of knowledge work.
In this perspectives paper, we introduce \ifg[full] as a natural evolution from Information Foraging. Drawing on current research, we illustrate how user behaviours are changing as foragers become farmers, and discuss the implications of Information Farming for future practices, systems, and research.

\section{Background}\label{lbl_background}
Hunter–gatherer societies depended on naturally occurring food sources in environments where resources were often scarce and unevenly distributed. 
Their foraging strategies were dynamic, continually adapting to the availability and distribution of resources.
Drawing on these parallels, the Berry Picking Model~\cite{Bates1989TheInterface} and Information Foraging Theory~\cite{Pirolli1999}, 
introduced ecological metaphors that became dominant conceptual framings of information seeking and retrieval behavior.
These models depict users as navigating across multiple information patches, opportunistically collecting relevant pieces while their information needs evolve dynamically during the search process.
While these assumptions have proven powerful for understanding search in environments dominated by externally available information, they become less adequate when users can proactively generate, shape, and refine information within the interaction itself.
%
%
To motivate this shift, we first revisit Berry Picking and Information Foraging Theory, then draw on the Neolithic Revolution as a historical analogy to illustrate how changes in resource acquisition can drive broader transformations. This context grounds our argument for a transition from Information Foraging to Information Farming.



.



\vspace{-5mm}
\subsection{Berry Picking}

\begin{quote}
\textit{“The berries are scattered on the bushes; they do not come in bunches. One must pick them one at a time.”} — Bates, The design of browsing and berry-picking techniques for the online search interface~\cite{Bates1989TheInterface}
\end{quote}

Early \ir[full] models often assumed that users were attempting to satisfy a fixed information need. And, if an ideal query could be formed then it could be matched against a corpus to produce the perfect ranking of documents (the classic IR paradigm~\cite{Robertson1977Theories,Salton1988TermWeighting}).
Bates challenged this by observing that information-seeking behavior in practice was iterative and dynamic~\cite{Bates1989TheInterface}. 
Rather than searching for one definitive list of documents, users accumulated relevant information incrementally, collecting small, scattered pieces of information from multiple sources over time.

\begin{figure}[h!] 
    \centering
  \includegraphics[width=0.95\linewidth]{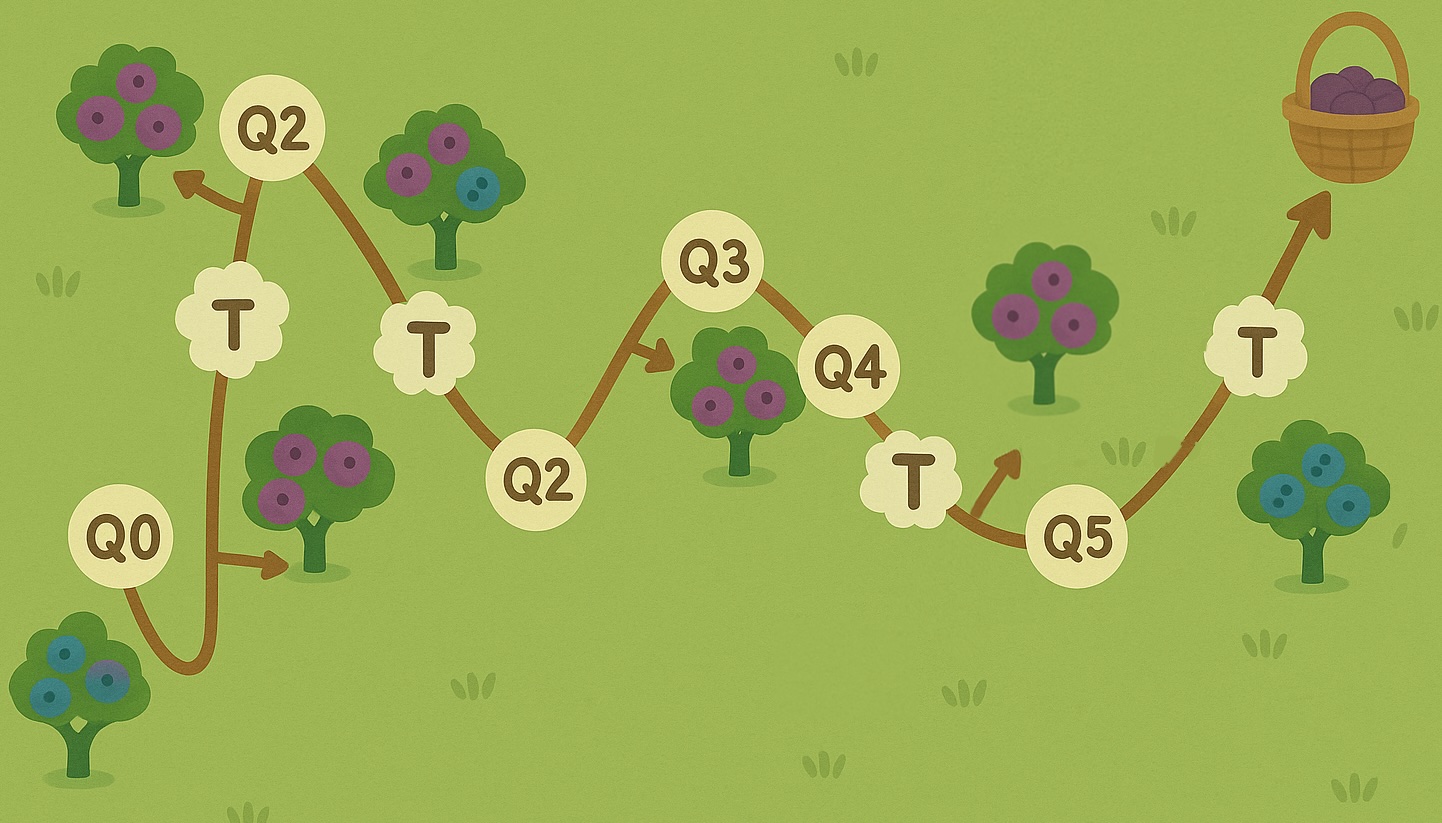} 
    \caption{The Berry Picking Model where the path users take through the information space evolves as they encounter new information (given their queries Q and their thoughts T collecting berries along the way until the end -- the basket of berries).}
    \Description{The Berry Picking Model where a users search evolves as they encounter new information.}
    \label{fig:berry-picking} 
\end{figure}

As shown in the illustration in Figure~\ref{fig:berry-picking}, Bates likened this process to a berry picker, moving from patch to patch, collecting berries (or ``bit a at time'' retrieval). 
As users encounter new pieces of information, their understanding of the problem space changes, their cognitive state changes, and their search strategies change as observed through different user behaviours, like query refinement or changing the sources being searched. 
The search process, therefore, is a sequence of micro-decisions, in which each retrieved piece of information (e.g., document, snippet, citation, image)
can alter the trajectory of exploration and the path taken. 
This paradigm highlighted the non-linear, opportunistic, and adaptive nature of information seeking, situating it closer to real-world practices such as browsing, chaining, and serendipitous discovery~\cite{CRL13386,Ellis1989}. 

\subsection{Information Foraging Theory}
\begin{quote}
\textit{“We should expect adaptive systems to evolve toward states that maximize gains of valuable information per unit cost.”}
— Pirolli \& Card, Information Foraging Theory~\cite{Pirolli1999}
\end{quote}


Information Foraging Theory~\cite{Pirolli1999} draws on Optimal Foraging Theory~\cite{MacArthurPianka1966Optimal} from behavioral ecology~\cite{KrebsDavies1997BehaviouralEcology}, which models how organisms make strategic decisions about exploiting food patches to maximize net energy gain. In this framework, foragers continuously weigh the benefits of staying in a patch against the costs of moving to a new one, optimizing their behavior in dynamic environments. 
Translated into the information domain, this perspective frames search as an adaptive optimization process in which, information seekers, conceptualized as ``\textit{informavores}''~\cite{Miller1983Informavores}, who balance the costs of searching, navigating, and processing information against the expected gain of the information obtained~\cite{azzopardi2011economics}. 

\ift~is comprised of three interrelated models: scent, patches, and diet. 
The \textit{information scent} model captures how perceptual cues, such as titles, highlighted, and other visual signals, indicate the potential relevance or value of a source. Strong scent increases the likelihood that a user will invest effort in exploring a resource, while weak scent may prompt abandonment. 
The patch model complements this by describing how seekers allocate time and effort within and across information sources.
Drawing on the \textit{Marginal Value Theorem}~\cite{Charnov1976Optimal}, it predicts that users will leave a patch (e.g., website, database) once the expected gain from continued search falls below the average gain available elsewhere.
Finally, the diet model formalizes decisions about which information item types to consume according to their profitability and selecting the subset that maximizes the overall rate of information gain. 
This translates to decisions about which types of information to pursue: a seeker may focus on a narrow set of high-value sources (specialist strategy) or include a broader range of sources to maximize overall information gain (generalist strategy), depending on the expected utility relative to cost.
Together, these three constructs provide the formal framework for understanding information seeking as a process of continuous trade-offs among effort, time, and utility.

The connection between \bpg[full] and \ift[full] is both conceptual and historical. \bpg[full] emphasizes the micro-level, experiential process of gathering information in small, evolving increments, while \ift[full] formalizes these behaviours within a cost–benefit framework at the macro level. 
This synthesis, which captures both the cognitive dynamics of evolving information needs and ecological constraints in search environments, has proven influential;
informing system design principles that support iterative search, dynamic query reformulation, and navigation across heterogeneous information landscapes. 
These paradigms have stood the test of time, despite improvements and enrichment to information access of the past decades (e.g., knowledge cards, instant answers, federated search, digital libraries). 
Moreover they have been shown empirically to explain many different search behaviours in various settings, for example, web browsing patterns~\citep{Huberman1998,Mulholland2008}, modern query reformulation strategies~\citep{White2009}, professional search ~\cite{Dwairy2011}, exploratory search~\cite{Woodruff2012} search as learning~\cite{Barack2023Visuospatial}, and more.
However, with the advent of \genai, information behaviours are changing, and a new paradigm is emerging.

\subsection{From Foraging to Farming}

\begin{quote}
\textit{``From the appearance of the human race some 7 million years ago, until the introduction of agriculture, hunting and gathering was the only food procurement strategy practised. The transition to agriculture, which led to the rise of civilisation as we know it, has, therefore, rightfully been termed the Neolithic Revolution.''} --  Wesidorf, Foraging to Farming~\cite{Weisdorf2005ForagingToFarming}
\end{quote}

Access to essential resources such as food and water is critical for survival.
Different species, including early humans, evolved foraging strategies adapted to the availability and distribution of these resources. 
For hunter-gatherer societies, survival depended on accumulated knowledge, passed through generations, about where and how to obtain resources and how to avoid over-exploitation.

During the ``\textit{Neolithic Revolution}'', around 10,000 to 12,000 years ago, a fundamental shift occurred which saw humans transition from primarily foraging-based societies to agrarian ones~\cite{bellwood2005firstfarmersorigins}.
The changes to how societies could obtain food resources, through the adoption of farming, also led to profound changes in society itself, by fundamentally changing social structures, settlement patterns, and people's relationship with sustenance and the environment.
While, initial efforts in farming yielded mixed results, over time, more systematic and reliable means of cultivation and production were developed such as deliberate planting schedules, selective breeding of flora and fauna, and advanced harvesting mechanisms. This led to higher yields and increased food security~\cite{barker2006agriculturalrevolutionprehistory,zohary2012domesticationplantsoldworld,hodder2006leopardstaleCatalhoyuk,wright1994groundstonetoolshuntergatherer,kuijt2000lifeneolithicfarmingcommunities}. As a consequence, hunting and gathering activities decreased, and farming and cultivating activities increased. 
This shift was not swift, but instead it was gradual and uneven, unfolding over hundreds of years, as early farming was labour intensive and often yielded lower nutritional diversity~\cite{Harlan1992CropsAndMan}.
Nowadays, farming has become so systemized and productive, most people are able to direct their efforts elsewhere~\cite{Smith1776WealthOfNations}.

The transition from migratory foragers to agrarian communities resulted in many changes to human society.
While most changes could be seen as positive (e.g.,  greater access to food and food security, support for non-producing specialists--like artisans and craftspeople~\cite{bellwood2005firstfarmersorigins, zohary2012domesticationplantsoldworld,robinson1982agriculturalsurplus,marx1867primitiveaccumulation,FAO2011surplusprinciples,diamond1997gunsgermssteel,hodder2006leopardstaleCatalhoyuk}), some changes were neutral (e.g., writing, record-keeping and taxation systems~\cite{diamond1997gunsgermssteel, fuller2007contrastingpatternscropdomestication}) and others were perceived as negative (e.g., sedentarism, dietary shifts, mono-cultures, elite social classes~\cite{marx1867primitiveaccumulation,barker2006agriculturalrevolutionprehistory, kuijt2000lifeneolithicfarmingcommunities,bellwood2005firstfarmersorigins}). 
Nonetheless, the agricultural surplus, enabled by farming, was fundamental in the transformation from small, egalitarian communities to large, complex societies defined by occupational specialization, stratified hierarchies, and institutional frameworks~\cite{diamond1997gunsgermssteel, barker2006agriculturalrevolutionprehistory, kuijt2000lifeneolithicfarmingcommunities}. 
While the \genai revolution is in its infancy, the parallels to the Neolithic revolution are becoming increasingly apparent.
In this paper, we explore how the current revolution is evolving people's information-interactions from foraging to farming.

\section{Information Farming}\label{lbl_farming}
\begin{quote}
\textit{“The first revolution that transformed human economy gave man control over his own food supply. Man began to plant, cultivate, and improve by selection edible grasses, roots, and trees.”} — Childe, Man Makes Himself~\cite{Childe1936ManMakesHimself}
\end{quote}

\begin{figure}[t] 
    \centering
    \includegraphics[width=0.9\linewidth]{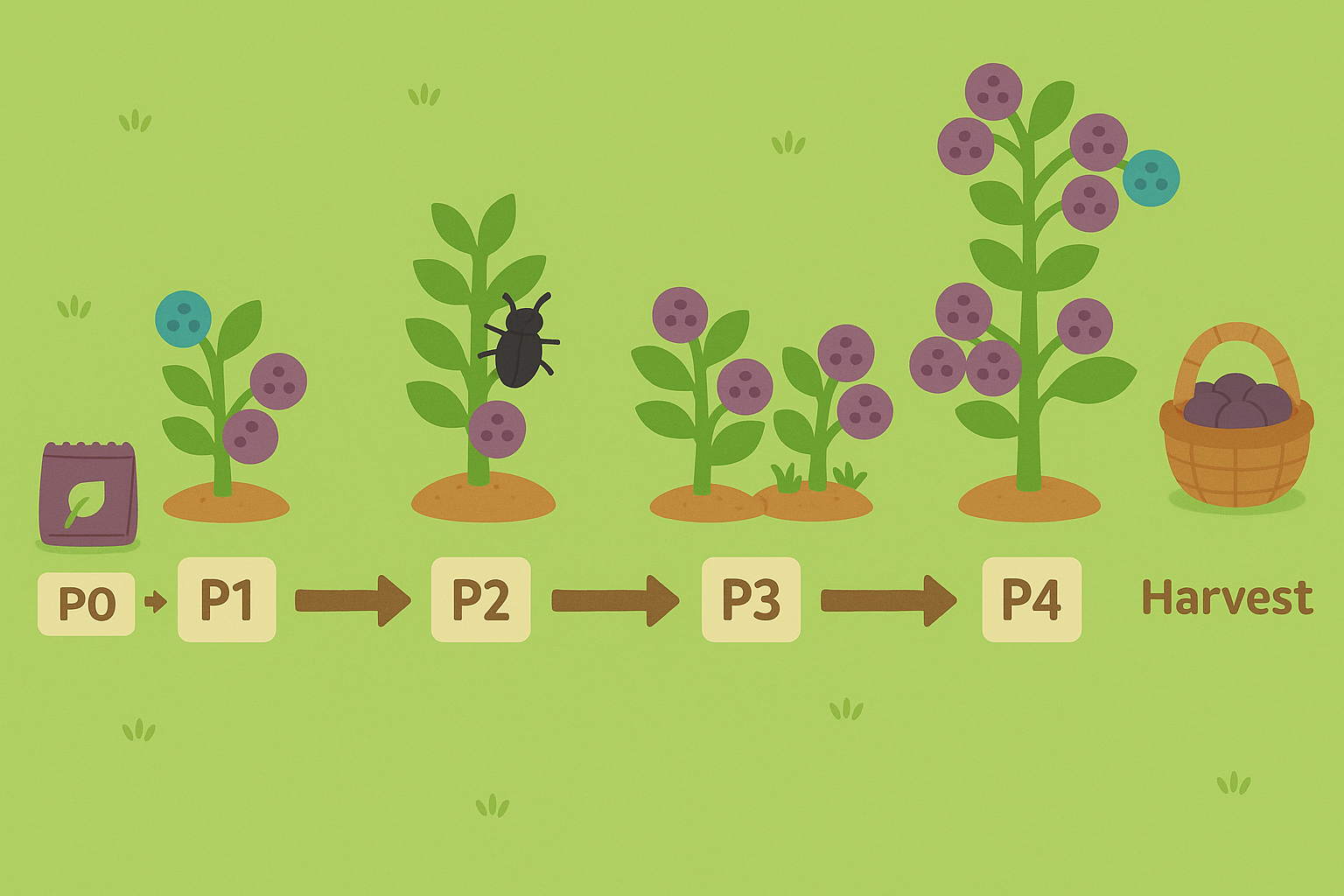} 
    \caption{The Berry Growing Model where the user plants an initial seed prompt (P0) and then cultivates their crop through iterations: growing (P1), removing pests (P2), weeding (P3), and other activities until harvest (P4).}
    \Description{The Berry Growing Model.}
    \label{fig:berry-growing} 
    \vspace{-6mm}
\end{figure}

The emergence of \genai technologies (e.g., \llm, \rag, Agentic IR systems) are fundamentally changing how people interact with information. 
Rather than solely foraging, people are beginning to move towards farming. 
This natural evolution has arisen because \genai provides an information rich medium (``\textit{a plot}'') in which people can ``\textit{sow},'' ``\textit{grow},'' and ``\textit{harvest}''  the information that they desire. 
Much like how farming allows humans to control food production, information farming can provide greater control over the information produced and subsequently consumed (subject to the growth medium and farming systems in place). 
Farming information allows users to produce the desired information, in a format that they prefer, resulting in higher yields (gains) without the need to forage from patch to patch. 

\vspace{5mm}
\begin{table*}[th]
\centering
\caption{Analogy between agricultural practices and Information Interactions with \genai.}
\vspace{-2mm}
\label{tab:genai-farming-analogy}
\small
\begin{tabular}{p{2.4cm} p{8cm} p{5.5cm}}
\toprule
\textbf{User Interactions} & \textbf{Description} & \textbf{Example Action} \\
\midrule
\textbf{Seeding} & Just as planting seeds determines the crop’s potential, crafting an initial prompt seeds what information will be produced. & ``\textit{Make a list of laws relevant to information retrieval and seeking.}'' \\
\textbf{Growing} & Like a sprouting plant, this phase involves further prompting to grow more branches and berries through subsequent interactions. & ``\textit{For each law, provide a detailed description, including citations and examples.}''\\
\textbf{Pruning} & Pruning removes excess or tangential growth to focus on what's essential. Similarly, asking for refinements of the information involves trimming irrelevant or repetitive content to sharpen the focus on core concepts. &
``\textit{Shorten the description for each law, so that we have one sentence defining the law, one sentence to explain its application, and one sentence provide a key example.}''\\
\textbf{Weeding} & Weeding removes unwanted and harmful outputs in the plot. This means identifying and eliminating hallucinated text or citations, inaccuracies, biased statements, and so on. & ``\textit{The math for Bradford's law is not correct it should be: $1 : n : n^2$, and Murphy's law, is not relevant, remove it.}''\\
\textbf{Fertilizing } & Fertilizing enriches growth; in \genai, this corresponds to contextual enrichment through added data or retrieval-augmented generation to enhance reliability or direct the growth. & ``\textit{Here is a paper~\cite{azzopardi2009query} on the Least Effort In Finding aka Leif's law. Include that in the list too.}''\\
\textbf{Breeding} & Comparing competing outputs and selecting the best, can lead to improved outputs given the criteria set.
  & \textit{``Phrase the law on fairest in ten different ways, then compare pairs of phrasings, in a round robin competition to select the best phrasing.''}    \\
\textbf{Cross-pollinating} & Cross-pollination introduces diversity and resilience into the outputs. By combining different ideas, outputs and sources together can give rise to new and novel outputs.  & \textit{``Take Newton’s laws of motion and reinterpret each one in the context of \ir''}. \\
\textbf{Harvesting} & Harvesting refers to extracting usable results like answers, lists, timelines, or structured summaries from the outputs. & \textit{``Take all the laws we have discussed and presented them in order of appearance in the literature.''} \\
\textbf{Packaging} & Packaging refers to refining and formatting  outputs (like bailing up the wheat) for subsequent processing or tasks. & \textit{``Create a \LaTeX~ table with two columns for law and description along with the relevant citations.''}\\
\textbf{Seed-Making } & Like saving seeds, researchers create reusable meta-prompts that capture refined workflows, ensuring consistency and efficiency in future projects. &  ``\textit{Review the conversation so far and design an improved prompt that would reliably reproduce a summary of what we discussed.''} \\
\textbf{Re-planting } & Re-planting symbolizes the reuse and adaptation of successful prompts in new contexts or research domains. & ''\textit{Create a table contain the list of laws in Information Retrieval in chronological order, provide the law, its description and citations, in \LaTeX.}'' \\
\textbf{Composting} & Composting converts waste into nourishment. Similarly, reusing discarded or partial AI outputs and user notes can be enhanced and refined, turning ``\textit{waste}'' into innovation. & ``\textit{Take these notes on the laws of \ir: [insert notes], and organize them in a coherent summarize, fill in any details where appropriate.}''\\
\bottomrule
\end{tabular}
\vspace{-1mm}
\end{table*}

\vspace{2mm}\noindent{\bf Plot Selection}
Rather than visiting different patches, information farmers choose between and among different plots. 
Just as soil and land preparation is fundamental in traditional farming for laying a fertile foundation for growing, in the context of information-interactions, different foundational models (e.g., LLama~\cite{touvron2023llamaopenefficientfoundation}, Mixstral~\cite{jiang2024mixtralexperts}, DeepSeek~\cite{deepseekai2025deepseekv3technicalreport}), RAGs (e.g., Google, Bing), Agents (e.g., Copilot, ChatGPT, Claude, Gemini) or tools (e.g., Grammarly\footnote{\small{\url{https://www.grammarly.com}}}, PaperPal\footnote{\small{\url{https://paperpal.com/}}}, Writefull\footnote{\small{\url{https://www.writefull.com}}}) are selected based on which is most likely to lead to the greatest yields (subject to cost, access and availability).

\vspace{2mm}\noindent{\bf Information Seeding and Breeding}
Once a medium is selected, farmers undertake an array of information-interactions (see Table
~\ref{tab:genai-farming-analogy}) to farm information.
Rather than foraging, a farmer plants a ``\textit{seed}'' in the form of a prompt, and the \genai system, acting as the farming infrastructure, ``\textit{grows}'' information in the farmer's ``\textit{plot}'' (see Figure~\ref{fig:berry-growing}).
Through subsequent interactions, farmers curate the shape and size of the information grown: whether this is to: ``\textit{grow}'' additional information or more detailed information, ``\textit{prune}'' and trim down the information grown,  or ``\textit{weed}'' and root out any errors, hallucinations or biases within the information grown. 
These interactions are distinctly different from foraging, where forager needs to visit different patches in the hope of finding berries, which they then have to pick, gather and process themselves. 
Moreover the farmer can perform actions to enrich the information produced by ``\textit{fertilizing}'' and enriching the context with external information (or instruct an agent to do so on their behalf). And, furthermore, farmers can go beyond what exists in the wild, and generate new information, by ``\textit{breeding}'' and/or ``\textit{cross-pollinating}'' different ideas, pieces of information, and/or, sources together to create new information (i.e., new kinds of berries) that did not exist previously.

\vspace{2mm}\noindent{\bf Information Harvesting}
Unlike the forager, the farmer has more control over the ``\textit{packaging}'' and ``\textit{processing}'' of how the berries by prescribing how the information should be formatted, structured, and presented.
The forager, in contrast, must first pick the berries, and then process it manually following collection. 
The cost of foraging and processing, therefore, is likely to be higher than the cost of farming. 
Additionally, yields from farming are likely to be much higher than foraging. 
Taken together this suggests that farming practices will increasingly be adopted, especially as systems and infrastructure to support information farming are built and refined. 

\vspace{2mm}\noindent{\bf Information Cultivation}
In addition to these growing and harvesting oriented interactions, farmers may want to preserve or store their information-interactions by creating a synthesis in the form of a prompt.
This is akin to a farmer collecting the seeds from the plants grown, so that they can plant them again, and bear similar fruit in the future. 
The new prompt (seed) essentially allows the information farmer to regenerate cultivated information. These seeds can  be planted in another medium (e.g., a different \genai agent or \llm) to produce similar but different information. 

Just as crop rotation in agriculture ensures soil fertility and biodiversity, and avoids nutrient depletion. Information farmers can rotate between mediums, combine and mix yields together, as a way to mitigate risks of bias, repetition, or overfitting. 
For example, after relying heavily on one agent to ``\textit{create a list of the laws of IR},'' the information farmer might rotate to another agent, to see what is produced in another plot, and then combine the harvests together.  
While real-time information-interactions may result in bountiful yields of relevant information, it may not yield all the relevant information. 
However, Agentic IR solutions enable the information farmer to sow the seeds, and then wait for the agent to grow information over time; without the farmer needing to continuously tend to the plot (e.g., researcher agents). 
In contrast to the forager, who needs to actively and continuously search for additional information (bearing an increasing cost per unit of subsequent information accrued). 
While the law of diminishing returns still holds for farming, the search and processing costs required to forage is not incurred, so the farmer can attend to other (more productive) activities.

\vspace{2mm}\noindent{\bf Farming Protections}
More advanced farming systems have evolved over time to help farmers guard against pests and bugs, and other types of crop failures. 
Analogously, information farmers need to be wary of and guard against harmful intrusions such as misinformation, adversarial, or toxic outputs that degrade and destroy the quality of the harvest. 
As such farmers can adopt methods (e.g., weeding) or tools and workflows that reduce hallucinations (e.g., RAG, Agentic IR, fact checkers) and/or select plots where steps have been take to sanitize or decontaminate the mediums (e.g., that has undergone appropriate alignment~\cite{kaufmann2024surveyreinforcementlearninghuman}, training data selection~\cite{liu2025proactive_defense}, other safeguards~\cite{shi2024llm_safety,huang2023survey_safety}).
Other protections may come in the form of fences that guard against unwanted intrusions (e.g. jail breaking, prompt injection)~\cite{fu-etal-2024-vulnerabilities}. In the \genai era, setting boundaries through ethical constraints, privacy controls, and access permissions is essential to prevent misuse or leakage of sensitive data. 
For example, a research team may confine their AI-assisted workflows within secure sandboxes, ensuring data security and privacy~\cite{dong-etal-2024-attacks}.
Such controls over the environment are often not possible when foraging in the wild.

\vspace{1mm}
In sum \ifg[full] is an evolution in information interaction brought about by \genai technologies. 
Through deliberate plot selection, seeding, cultivating, and harvesting, users leverage \genai and related systems to control and increase information yield, quality, and reproducibility.

\subsection{Foraging vs. Farming}
Foraging and farming represent two constitute complementary practices for encountering, retrieval, acquisition, creation, and production of information. The shift towards farming practices goes beyond efficiency to encompass cultivation over extraction.
To clarify this distinction, Table~\ref{tab:farming-foraging-comparison} summarizes the key differences between information foraging and information farming.
Information Foraging characterizes information seeking as an environment-driven process in which users adapt their strategies to the availability and structure of existing information, with limited control over outputs.
Foragers must visit many distributed and heterogeneous sources, investing substantial time and effort to transform what they find into a usable or desired output.
For example, consider a researcher conducting a literature review on the various laws of Information Retrieval. 
Under a foraging approach, this task requires exploring multiple databases, issuing numerous queries, following citation trails, and incrementally gathering relevant papers and proposals.
This process entails sustained effort across multiple patches to collect the necessary references and background material, which must then be further processed: read, extracted, summarized, and synthesized (as described by the Information Search Process~\cite{Kuhlthau1991}).
The amount of information ultimately found is therefore an function of the effort required, the rate of gain, and the time needed to process and handle information, subject to practical constraints such as deadlines, attention, and available resources~\cite{azzopardi2014economics}.
In systematic reviews, for instance, the number of items that can be processed is directly constrained by the available budget or time, often leading to a narrower scope to reduce the volume of retrieved material~\cite{yang2023gptqueries,scells2020automatic}.

In contrast, Information Farming describes an evolution of information interaction in which users act as active cultivators, deliberately selecting a plot (e.g., a \genai tool or system) suited to their goals and constraints.
Through intentional seeding (prompt design), iterative cultivation, and refinement, users shape both the quality and quantity of the information produced. 
Rather than passively accepting what is found, farmers actively guide growth by pruning errors, enriching content, and guarding against harmful or low-quality inputs.
Returning to the literature review example, a researcher may use a \genai tool to identify foundational works, generate summaries, and structure these into draft annotated bibliography. They can then iteratively explore related concepts, potentially across multiple tools, and feed the resulting material back in to refine and expand the review. 
In this way, the researcher retains control of the direction and quality of the output, with tools supporting sustained information growth rather than isolated retrieval events.
With Agentic IR systems(e.g., DeepResearcher~\cite{zheng2025deepresearcherscalingdeepresearch}, Perplexity AI~\footnote{\url{https://www.perplexity.ai/}}, Grok’s DeepSearch~\footnote{\url{https://grok.com/}})
such cultivation can occur at scale and with reduced need for constant human supervision, allowing effort to be redirected to higher-level reasoning and synthesis.
Compared to the high costs incurred during foraging, where effort is dominated by searching, navigating, and post-processing found information, farming concentrates effort on the design and management of growth conditions. This enables higher-quality, more tailored yields, and lowers barriers for users who may otherwise struggle with complex search tasks (e.g., non-experts conducting literature reviews).

Finally, Information Farming supports the preservation and reuse of seeds for future (re-)planting, enabling reproducibility and portability across tools, contexts, and agents.\footnote{\small{Shifting the dynamic from ``stuff I've seen'''~\cite{dumais2003sis} to ``stuff I've grown.''}}
While both paradigms share the overarching objective of maximizing information gain per unit of effort, they diverge in their assumptions, strategies, and outcomes. 
Information Foraging optimizes within a given environment with the objective of efficiency under scarcity (i.e., it is constrained by the availability and accessibility of existing information).
Information Farming, in contrast, optimizes through environmental design with the goal of enhancing yield, sustainability, and resilience through controlled cultivation of ecosystems for knowledge production.
Put another way, the forager adapts to what is found, the farmer shapes what is grown. 
We note that foraging and farming represent two complementary strategies, where information found through foraging can be refined through farming, and farmed information can enable and encourage foraging. 
\vspace{6mm}
\begin{table*}[t]
\centering
\caption{High-level comparison of Information Foraging and Information Farming.}
\label{tab:farming-foraging-comparison}
\vspace{-2mm}
\small
\begin{tabular}{p{2.6cm}p{6.8cm}p{6.8cm}}
\toprule
\textbf{Dimension} & \textbf{Information Foraging} & \textbf{Information Farming} \\
\midrule
\textbf{Primary Metaphor} & Searching and gathering from scattered patches (berry picking and foraging from Behavioural Ecology) & Planting, cultivating, and harvesting in selected plots (farming from Agricultural studies) \\
\textbf{Interaction Style} & Opportunistic, exploratory, based on availability & Deliberate, iterative, tailored; from seed to harvest \\
\textbf{Control over Output} & Limited; relies on what is found & High; information is shaped, pruned, and bred for quality and format \\
\textbf{Medium} & Uncontrolled, wild, diverse online and offline sources & Prepared, Pre-trained, curated LLMs, RAGs, \& \genai\\
\textbf{Effort Distribution} & High cost in search and manual post-processing & Effort in prompt design and iterations to cultivate, breed and harvest \\
\textbf{Yield} & Variable; depending on patch information density & Higher and more predictable through controlled growth and harvesting \\
\textbf{Innovation Potential} & Limited to recombining found information & Enables creation of new information through cross-pollination and synthesis \\
\textbf{Error Handling} & Manual filtering and correction after search & Active pruning (weeding), error and bias reduction during cultivation \\
\textbf{Temporal Dynamics} & Continuous searching and effort for each new need & Seeds (prompts) can regenerate similar outputs; portable across contexts \\
\textbf{Risk} & Variable, subject to user ability to avoid inaccurate/biased information & Controllable, possible to manage risks with safeguards (e.g., built-in fact-checkers) \\
\textbf{Scalability} & Constrained by user attention and energy & Scalable via AI agents and persistent, unattended growth \\
\textbf{Safeguards }& Exposed to hostile or unsafe sources, little containment & Boundaries set by ethical constraints, privacy controls, and responsible AI safety \\
\bottomrule
\end{tabular}
\end{table*}

\section{Information Farming in  Practice}

Recent studies spanning everyday web search, professional knowledge work, and search-as-learning settings reveal notable shifts in information behaviours and preferences when using \genai technologies.
In this section, we highlight how these practices can be conceptualized in light of \ifg[full]. It is worth noting that many of the observed actions are not mutually exclusive nor do we conduct an exhaustive survey.

\vspace{2mm}\noindent{\bf Seeding} Studies across various populations and tasks have seen that prompting generative systems is different to querying traditional search systems~\cite{spatharioti2025llmbasedsearch,arguello2023generativeir,kaiser2025chatgptgoogle,liang2025bingchat,park2025choicematessupportingunfamiliaronline,yen2025searchgen,liu2024gptsal,johnston2024student,wardle2025evolving}.
A consistent observation across these studies is for prompts to be longer than search queries, informal and conversational in tone, and more command/instruction oriented.
 As people interact more with \genai tools, and experiment through trial and error, their prompting styles evolve as they better understand how it affects outputs and the yield that is produced~\cite{wardle2025evolving,johnston2024student,lindrup2025promptmachine}. 
This is in contrast to foraging where work from the early 2000s found that longer search queries led to higher user satisfaction~\cite{belkin2003} and system effectiveness~\cite{kelly2005}, but, in practice, search queries remained relatively short because they are likely to yield a higher rate of gain~\cite{azzopardi2009query}.
 

\vspace{2mm}\noindent{\bf Growing, Pruning and Weeding}
Much like in farming, where growing, pruning, and weeding are part of the process, we also see instances of these behaviours when users engage with \genai.
In several studies, growing behaviours have been observed through the multi-turn conversations with \genai systems~\cite{arguello2023generativeir,liang2025bingchat,spatharioti2025llmbasedsearch,lee2025knowledgeworkers,liu2024gptsal,kim2025reliance,park2025choicematessupportingunfamiliaronline,yang2025searchchat,alam2025blingfatih}, where users focus on expanding and developing the topic of interest. 
When outputs become too verbose or go off topic,  users were observed to actively request outputs be pruned (e.g., summarized, shortened)~\cite{yang2025searchchat,kim2025reliance,venkit2025search,wardle2025evolving} to remove extraneous information and ensure that the conversations remain relevant to their needs. 
Due to the stochastic nature of \genai systems, and their tendency to hallucinate, users have been observed performing weeding behaviours. For instance, to eliminate inaccurate, wrong, or made-up citations, users asked for checks to occur or conducted manual checks to determine if citations existed~\cite{Peterson_2025,dai2025next,liu2024gptsal,venkit2025search}. While having to weed out inaccuracies degraded user trust, some users understood that this is a current limitation when using \genai~\cite{yang2025searchchat}. 

\vspace{2mm}\noindent{\bf Fertilizing and Cross-Pollination}
Ensuring that generated responses are on-topic, accurate, and relevant to the user's information need is critical for success. 
One way to do this is to fertilize the process with relevant information and the prototypical example being RAG as the system-oriented approach to accomplish
this. 
However, users have also been observed doing this by manually inserting different pieces of  information from a variety of sources, including traditional search engines as well as other generative systems, into the \genai tool to produce more desirable outputs~\cite{arguello2023generativeir, mayerhofer2025, yang2025searchchat} 
By combining information from different conversations or sources (e.g., external documents, model knowledge), users engage in cross-pollination to create new or more diverse outputs~\cite{yen2025searchgen}. In doing so, they draw upon the complementary nature of foraging and farming practices to curate and refine their results as part of their information seeking task (e.g., by including a library tutorial when asking for a function or using generative systems to suggest search terms~\cite{yang2025searchchat}). Users have also been observed performing composting actions, by re-using failed conversations and past outputs in subsequent sessions to produce new outputs~\cite{yen2025searchgen}.
Methods have also been proposed to conduct automated cross-pollination using agentic systems that embody different concepts, results, or user preferences that might better reflect varying user needs and requirements, and then using those different agents as part of the generative workflow~\cite{wu2024autogen,zhang2025agentrec,park2025choicematessupportingunfamiliaronline}.
The caveat is that manual fertilization and cross-pollination require the use of ancillary tools to help store, track, and transfer information due to a lack of built-in support~\cite{yen2025searchgen,park2025choicematessupportingunfamiliaronline}. This indicates that prior work on the importance of organizing information still remains relevant today~\cite{choi2021orgbox,li2023orgbox} and that work remains to make integration seamless.

\vspace{2mm}\noindent{\bf Re-Planting and Soil Preparation}
It is a known phenomenon that currently one's choice of model and prompt can non-trivially effect the outcome of the system. ~\citet{chowdhury2024} issued the same queries to six different generative systems and observed that at a high-level the results contained similar information but the actual outputs varied wildly. In solving programming tasks, participants were observed to have tried different tweaks to the same prompt to get the model to target new problems or incorporate new pieces of information (e.g., steps from a tutorial)~\cite{yen2025searchgen}. While the stochasticity of current systems makes them diverse and creative, it also limits their potential when two users issuing a similar prompt may produce dramatically different yields.

\vspace{2mm}\noindent{\bf Harvesting and Packaging}
In contrast to foraging, information farming can more readily facilitate the incremental harvest and production of information for consumption~\cite{yen2025searchgen,wardle2025evolving}. This is due, in part, to the fact that much of harvesting and packaging are tedious (but important) steps that can be time consuming for humans but more easily automated by generative systems (e.g., formatting information into a table).
The ease of harvesting and packaging information can be beneficial when users are knowledgeable in the area and are not solely relying on the capabilities of the system~\cite{liu2024gptsal,yang2025searchchat,mayerhofer2025}.  
In contrast, users understand that this easy information generation can have negative effects when users are less knowledgeable or take the outputs at face value~\cite{johnston2024student,kosmyna2025cognitive_debt}.
Moreover, there are also early indicators that over-reliance on these systems may lead to ``\textit{cognitive debt}'' (i.e., less retained knowledge) in habitual users~\cite{kosmyna2025cognitive_debt,lee2025knowledgeworkers} due to the appeal of seemingly correct information that can be easily packaged for distribution. 
Additionally, the need for proper attribution and careful use of harvested and packaged information is critical to avoid risks associated with the rampant, naive inclusion of generated documents in downstream systems (e.g., model collapse, proliferation of biased information)~\cite{Peterson_2025,garetto2025rgb}. Such concerns reinforce the need for ``information labels'' (e.g., human-generated, human-curated, \genai-only) that are analogous to modern food labels (e.g., organic, non-GMO)~\cite{fuhr2018labels}.

\vspace{2mm}\noindent{\bf Farm Protection}
To ensure that users grow a good crop of information, systems must ensure that users are able to trust the outputs of the system and, more importantly, that the system presents accurate information and does not reinforce or create new biases.
Studies have demonstrated that systems can and do produce outputs that reinforce implicit and explicit biases of users~\cite{sharma2024controversialnews,spatharioti2025llmbasedsearch,dejong2025conformity,venkit2025search}, which can be amplified by responses that appear to be authoritative or definitive~\cite{arguello2023generativeir,kim2025reliance,wardle2025evolving}.
In many ways, protection strategies are still nascent but have begun to be explored through better presentation of information~\cite{karunanayake2024} to more explicit control mechanisms~\cite{dai2025next} to additional data and model refinement~\cite{dai2024biasir} to the use of multiple, collaborating agents to improve diversity~\cite{wu2024autogen,zhang2025agentrec,park2025choicematessupportingunfamiliaronline}.

\section{Discussion}
\begin{quote}
\textit{"Why farm? Why give up the 20-hour work week and the fun of hunting in order to toil in the sun? Why work harder, for food less nutritious and a supply more capricious? Why invite famine, plague, pestilence and crowded living conditions?"} -- Harlan, Crops and Man, 1992\cite{Harlan1992CropsAndMan}
\end{quote}

\ifg[full]  is an emerging paradigm for interacting with information and \genai, bringing new opportunities alongside risks and challenges similar to those faced by early agrarian societies. This section provides an overview of those issues.

\vspace{2mm}\noindent{\bf Cultivation}
Central to \ifg[full] is the notion of deliberate cultivation, where users actively sow seeds and grow information  to produce rich, contextually relevant yields. 
The example of a researcher deploying \genai to systematically grow an annotated bibliography demonstrates this shift, resembling the intentional breeding and selection of crops that enabled farming societies to increase yields and crop complexity~\cite{Harlan1992CropsAndMan}. 
This proactive shaping of information and the information environment itself, contrasts with foraging assumptions where foragers passively gather scattered pieces of available information.
The benefits are evident in the form of higher and more sustained informational yields. And through iterative interactions with \genai tools, users can cultivate and harvest, not only dense (potentially) high‑quality knowledge, but also produce information that previously did not exist (e.g.,~\cite{kaiser2025chatgptgoogle,liu2024gptsal,yang2025searchchat,mayerhofer2025}). 
Yet unlike traditional agriculture bound by biological growth cycles, computational cultivation unfolds almost instantaneously, enabling the synthesis of entire fields of literature in moments, realizing, in informational form, the promise of abundance once envisioned through early agricultural innovation~\cite{Childe1936ManMakesHimself}.

\vspace{2mm}\noindent{\bf Control} 
The new array of user interactions afforded by \genai technologies give information farmers more control over the information cultivated and curated~\cite{arguello2023generativeir,yang2025searchchat,venkit2025search}.
Farming enables more precise specification of information needs over keyword search.  For instance, a farmer could state: ``\textit{Create a list of laws associated with IR, in chronological order. For each law provide a description of the law, cite who proposed the law. Then provide examples of how the law has been applied, citing relevant works.}''.
Whereas the forager would need first need to find each law and look for applications of each (via searching and browsing), before processing the retrieved content (assuming it exists).
Instead, under the farming paradigm the farmer engages with the \genai tooling to create and shape the desired information.
This control mimics that in agricultural farming, wherein more, higher quality food (information) can be attained with greater regularity and lower effort. 

\vspace{2mm}\noindent{\bf Specialization} 
Just as the transition to agrarian societies led to the development of specialized roles and jobs that were not always directly related to farming, \genai is likely to transform knowledge work. 
For example, there has been a rise of jobs that have ``\textit{prompt engineering}'' as a core skill and even as a job title~\cite{vu2025promptengineeranalyzingskill,zhang2023prompt,SPARKES202317}, which demand the use of new tooling to assist in finding and optimizing prompts.\footnote{Such as \small{\url{https://flowgpt.ai/}} and \small{\url{https://dspy.ai/}}.} 
Beyond this, new frameworks for communication between generative systems and existing services and infrastructure are commonplace\footnote{\small{\url{https://modelcontextprotocol.io/}}} as well as higher level collaboration frameworks to facilitate interactions between agents, systems, and humans~\cite{shah2025agentsnot,zhang2025agentrec,wu2024autogen}. 
Such developments motivate the need for knowledge engineering and knowledge automation roles, whereby the farming infrastructure is assembled to farm information to satisfy specific, dedicated information needs.
Following agricultural farming developments, such expertise and workflows will likely be encoded into tools that users can adopt to more effectively and efficiently grow, harvest, and process information, which creates a variety of new avenues for the community to explore.

With these benefits, however, comes increased up-skilling and cognitive demands~\cite{Daum2025Digitalization}. Understanding and mastering new techniques, tools, and workflows aligns with industrial-era agricultural labour transformations. This skill shift elevates users from information seekers to knowledge cultivators, underscoring the changing nature of knowledge work, while also creating new and different limitations and risks due to the reliance on generative systems~\cite{jelson2025empiricalstudyunderstandstudents}.

\vspace{2mm}\noindent{\bf Cognitive Costs and Biases}
The transition to farming comes with various cognitive trade-offs. 
Just as the Neolithic adoption of sedentarism brought reductions in the diversity of physical activities and perhaps diminished critical vigilance~\cite{Smith1776WealthOfNations,Childe1936ManMakesHimself}, information farming may risk a decline in exploratory search behaviours and critical thinking. 
Recent studies show reliance on \genai tools can accumulate ``\textit{cognitive debt}'' where users fail to critically engage with content~\cite{kosmyna2025cognitive_debt}. 
In this state, effortful mental processes essential for independent thinking, deep comprehension, and creative insight are replaced or deferred. 
While this outsourcing eases cognitive load in the short term, it incurs long-term costs—such as diminished critical inquiry, reduced creativity, and greater vulnerability to manipulation~\cite{kosmyna2025cognitive_debt}. 
When individuals produce and consume \genai output without evaluating its accuracy or relevance, they risk forfeiting ownership of their ideas and internalizing shallow or biased perspectives~\cite{kosmyna2025cognitive_debt,Krsmanović_Deek_2025,georgiou2025chatgptproduceslazythinkers}.
The shift toward Information Farming may also strengthen confirmation bias and ease the formation of filter bubbles~\cite{Pariser2011,white2013bias}. Greater user control over prompts and iterative refinement, combined with generative systems’ tendency to align with user intent, can lead to information outputs that increasingly mirror users’ existing beliefs, extending and potentially intensifying biases previously observed in search-based interaction~\cite{azzopardi2021cognitivebiases}. Moreover, reliance on specific \genai systems that re-enforce and confirm existing beliefs may compound these effects.
While some of this is due to issues with generative systems (e.g., training data (mis)use~\cite{saracevic2024,Zhong_2023}, bias and fairness~\cite{wu2025doesragintroduceunfairness,abolghasemi2025evaluationattributionbiasgeneratoraware}), many are simply extensions of current information problems that are exacerbated by the ease of accessing information via \genai systems.


\vspace{2mm}\noindent{\bf Information Health and Sustainability}
Research into regenerative farming stresses the vital role of stewardship and sustainable practices~\cite{Beacham2023Contextualising}, reinforcing the need for ongoing maintenance and varied information cultivation techniques to ensure long-term ecosystem health.
While users can employ both foraging and farming strategies over and across multiple sources and multiple GenAI
tools as a way to create more diversified crops (much like was done in the early agricultural era~\cite{Harlan1992CropsAndMan}), active stewardship is still imperative at both the user-level and system-level. Users need to be vigilant and prune non-relevant content, weed out hallucinated or inaccurate content; while system builders need to train and evaluate \genai tools and systems to ensure outputs are accurate and unbiased.
Management is required to maintain the health of the farming ecosystems and guard against degradation through the proliferation of misinformation, disinformation and bias~\cite{Peterson_2025,garetto2025rgb}. 
In doing so, the risk of poor information health may be mitigated, much like how care and consideration must be placed in our daily diets due to ready availability of ultra-processed foods.

The development of community and economic networks around information farming resembles the societal surpluses that enabled labour specialization and cultural flourishing in agrarian societies~\cite{Smith1776WealthOfNations,Childe1936ManMakesHimself}. Collaborative platforms (e.g., HuggingFace) function as digital commons  where information cultivation fosters collective intelligence and innovation. 
Yet, caution is warranted, over-reliance on specific platforms may result in ``\textit{crop failures}'' when misinformation spreads widely or users prefer ``fast food''-like information that is vacuous but addictive.\footnote{See \small{\url{https://www.forbes.com/sites/robertbtucker/2025/06/20/is-chatgpt-making-us-stupid/}}.} These phenomena parallel nutritional and environmental challenges of monoculture agriculture~\cite{diamond1997gunsgermssteel}, underscoring the risks of degraded informational diets~\cite{Pirolli1995InformationEnvironments}.

\vspace{2mm}\noindent{\bf Gains, Yield and Effort}
 In agricultural terms, gain is considered to be the total amount of crop harvested (i.e., tonnes of berries) and yield refers to how effectively and efficiently one can produce this gain (i.e., tonnes of berries per hectare)~\cite{fischer205yield}. 
In the information domain, gain under farming and foraging reflects the same concept of information gain, which may be represented through different evaluation mechanisms (e.g.,~\cite{pradeep2025,arabzadeh2024}\footnote{\small{In these works they refer to the small bits of valuable information as nuggets, rather than berries, but conceptually they are similar.}}). 
However, yield under information foraging is directly proportional to user effort, under information farming it is more complex. 
Information farmers can build upon previous interactions to increase the return on investment in the short term, re-use prompting techniques in the medium term, and build automated workflow in the long term to compound the investment resulting in greater yields. 
To write a new article on ``\textit{The Laws of Human Computer Interaction},'' our forager needs to repeat their process again, exerting much the same effort, going from patch to patch to find the required information, but our farmer can replant seeds or reuse a previous workflow.

Moreover, information farmers have more actions to choose from (Table~\ref{tab:genai-farming-analogy}) that may substantially change the yield. For example, pruning is likely a much lower cost than weeding, since one must spend time finding the ``\textit{weeds}'' (e.g., hallucinated or inaccurate information) rather than cutting erroneous or non-relevant information. 
Complicating this is that some costs are upfront and fixed, while others are variable and rely on user and system interaction. 
These factors suggest that different optimal interaction strategies are likely to manifest depending on the farmer, the task, and the tools. For certain tasks, some farming workflows may lead to  higher gains and information yields than others, while foraging may be more efficient for other tasks. Understanding these trade-offs will be key in deciding when to employ one strategy over another. 

Under Information Farming, yield goes beyond simple measures of efficiency and ties into how usable the grown information is for the intended purpose. While a forager must accept the information as it is presented (e.g., a technical explanation on a government website), a farmer can tailor their crop to produce a version that is more useful to them (whether that be simpler or with examples)~\cite{merker2025axioms,roegiest2024}. 
In doing so, yield refers to not just how efficiently the information can be grown but how effectively the farmer can make use of it. The result is that evaluation of grown information is not just a matter of what information is presented but how fit for purpose it is.
Moreover, this process may result in growing previously non-existent information. While this information may be technically ``\textit{hallucinated},'' we argue that if it is relevant, desired, and accurate then this new information creates value (a ``\textit{compounded gain}'' so to speak).
Measuring such gains was not applicable in the context of foraging, but farming invites the measurement and evaluation of what is created (not just what is found or re-generated).

\vspace{2mm}\noindent{\bf Societal Dynamics and Inequalities}
Further, the ecological and social complexities of \ifg[full] demand careful attention to equitable access, control over \genai systems, reliability of outputs, and ethical governance, echoing the environmental and social tensions introduced by historical agricultural development~\cite{Harlan1992CropsAndMan,marx1867primitiveaccumulation}.
Just as the Neolithic Revolution gave rise to new social dynamics, occupational specializations, and community norms, \ifg[full] similarly reshapes how users collaborate and interact within information ecosystems.
It empowers individuals and communities to actively cultivate and co-create knowledge, fostering new forms of expertise and shared stewardship.
As systems enable cross-pollination between users and communities, new challenges emerge around collaborative information farming, such as fair attribution to users and systems, long-term maintenance, and the collective ownership of curated knowledge. 
This goes beyond current debates on copyright and data laundering involving \genai. There is also the critical risk of inadvertently reproducing exploitative dynamics akin to feudalism or sharecropping, where platform owners disproportionately benefit from farmed information.
Careful design is needed to ensure equitable access, inclusive participation, and sustainable governance, so that the benefits of information farming can be justly distributed across society, rather than concentrated among a few platforms.

Together, these dynamics suggest that \ifg[full] is not merely a technological change but a societal transition. One that will redefine information production, engagement, and stewardship in the digital era, demanding new tools, systems, methods and evaluation methodologies. This perspectives paper provides a starting point for further discussions and motivates many new research questions as this transitionary period unfolds.

\section{Summary and Conclusion}
\vspace{3mm}
\begin{quote}
\textit{“When by the improvement and cultivation of land the labour of one family can provide food for two, the labour of half the society becomes sufficient to provide food for the whole. The other half, therefore, or at least the greater part of them, can be employed in providing other things, or in satisfying the other wants and fancies of mankind”} — Smith, The Wealth of Nations (1776)~\cite{Smith1776WealthOfNations}.
\end{quote}

\vspace{3mm}
Bates’ \textit{\bpg[full]}~\cite{Bates1989TheInterface} and Pirolli \& Card’s \textit{\ift[full]}~\cite{Pirolli1999},
positioned information seekers as hunter-gatherers that moved between patches, collecting pieces of information from disparate sources, and adapting strategies to maximize their information gain and minimizing effort. 
Just as early humans transitioned from hunter-gatherers to agrarian communities during the Neolithic revolution, we have argued that users (along with systems) are undergoing a similar transition with the advent and adoption of \genai technologies. Continuing the metaphor from Information Foraging, we arrived at \ifg[full] which redefines this phenomena as a generative, co-evolutionary loop where query, system, and output continuously develop together. 
Here, the user’s role transforms from a forager to a farmer, actively shaping and refining outputs through interactive and iterative engagement with information using \genai systems.

First, \ifg[full] prioritizes \textit{cultivation over extraction} by empowering users to nurture and transform outputs rather than encountering and retrieving fixed and existing resources.
Second, it emphasizes \textit{control over adaptation}, empowering users to actively shape and refine information outputs, rather than continually adjusting to the constraints and unpredictability of what is encountered during foraging.
Third, it fosters \textit{specialization over generalization}, encouraging the emergence of distinct roles, expertise, and dedicated tools that optimize the processes of cultivation and harvest; whereas foraging relies on broad, adaptable skills and undifferentiated strategies for locating and gathering resources.

By positioning users as information farmers, users are situated as active participants that grow information and influence outcomes through an iterative process of prompting and feedback. 
\ifg[full] motivates processes for on-going maintenance, correction, and curation to promote trustworthiness and provides a common basis for mitigating biases and misinformation. 
Once information is grown, traditional practices for the harvesting, packaging, and distribution take place to provide the insights to the larger community. This can be accomplished through the distribution of concrete knowledge artifacts, prompts, or other tools to facilitate reuse and new growth. 

In conclusion, the transition from Information Foraging to Information Farming represents not merely a technological development but a fundamental shift in how people engage and interact with information: it empowers users to shape, cultivate, create and regenerate information.
 And as \genai technologies evolve, the tools and practices of information farming promise richer, more creative, and sustainable knowledge ecosystems, inviting researchers and users alike to help grow what comes next.

\newpage
\begin{acks}
The authors would like to thank our reviewers for their insightful and engaging feedback on this paper. 
\end{acks}
\balance
\bibliographystyle{ACM-Reference-Format}
\bibliography{references,transition-refs, behaviors,more-references,discussion-refs}

\end{document}